\documentclass[twocolumn,amsmath,amssymb,10pt,aps]{revtex4}
  
\usepackage[utf8]{inputenc}
\usepackage[T1]{fontenc}

\usepackage{amsmath}
\usepackage{amsfonts}
\usepackage{graphicx}
\usepackage{yfonts}
\usepackage{color}
\usepackage[normalem]{ulem}
\usepackage{amsthm}
\usepackage{bm}
\usepackage{bbm}
\usepackage{mathtools}
\usepackage{array}
\usepackage{placeins}
\usepackage{enumitem}

\newcommand{\der}{\mathrm{d}}

\def\<{\langle}
\def\>{\rangle}

\def\oper{{\mathchoice{\rm 1\mskip-4mu l}{\rm 1\mskip-4mu l}
{\rm 1\mskip-4.5mu l}{\rm 1\mskip-5mu l}}}
\DeclareMathAlphabet\mathbfcal{OMS}{cmsy}{b}{n}

\newtheorem{Example}{Example}

\begin{document}

\title{Quantum evolution with a large number of negative decoherence rates}

\author{Katarzyna Siudzi{\'n}ska and Dariusz Chru\'{s}ci\'{n}ski}
\affiliation{Institute of Physics, Faculty of Physics, Astronomy and Informatics \\  Nicolaus Copernicus University, Grudzi\k{a}dzka 5/7, 87--100 Toru{\'n}, Poland}

\begin{abstract}
Non-Markovian effects in quantum evolution appear when the system is strongly coupled to the environment and interacts with it for long periods of time. To include memory effects in the master equations, one usually incorporates time-local generators or memory kernels. However, it turns out that non-Markovian evolution with eternally negative decoherence rates arises from a simple mixture of Markovian semigroups. Moreover, one can have as many as $(d-1)^2$ always negative rates out of $d^2-1$ total, and the quantum evolution is still legitimate.
\end{abstract}

\flushbottom

\maketitle

\thispagestyle{empty}

\section{Introduction}

In quantum mechanics, it is impossible to completely isolate a physical system from the influence of environment. Therefore, the theory of open quantum systems is undergoing a rapid development \cite{RivasHuelga,Weiss}. Quantum processes that describe the evolution of open quantum systems are represented by time-parametrized collections of quantum channels $\Lambda(t)$, which are completely positive, trace-preserving (CPTP) maps \cite{BreuerPetr}. The family $\{\Lambda(t)|t\geq 0\}$ is referred to as a {\it quantum dynamical map}. It transforms an arbitraty initial state $\rho$ into an evolved state $\rho(t)=\Lambda(t)[\rho]$.

If the coupling between the system and the environment is relatively weak, it is justified to apply the Born-Markov approximation to the evolution equation. Then, the dynamics of the system is given by the Markovian master equation
\begin{equation}
\dot{\Lambda}(t)=\mathcal{L}\Lambda(t)
\end{equation}
with the Gorini-Kossakowski-Sudarshan-Lindblad (GKSL) generator \cite{GKS,L}
\begin{equation}\label{GKSL}
\mathcal{L}[\rho]= -i\hbar [H,\rho] + \sum_{\alpha}\gamma_\alpha \left(V_\alpha\rho V_\alpha^\dagger -\frac 12\{V_\alpha^\dagger V_\alpha,\rho\}\right).
\end{equation}
The generator is characterized by a Hamiltonian $H$, noise operators $V_\alpha$, and positive decoherence rates $\gamma_\alpha$. In the case of strong system-envoronment interactions, memory effects come into place. Going beyond the Markovian semigroup, one usually considers the master equation with a time-dependent generator $\mathcal{L}(t)$,
which has exactly the same form as $\mathcal{L}$ in eq. (\ref{GKSL}), but now $\gamma_\alpha(t)$, $V_\alpha(t)$ and $H(t)$ are time-dependent. Its formal solution is the map
\begin{equation}\label{T-exp}
\Lambda(t) = \mathcal{T} \exp\left( \int_0^t \mathcal{L}_\tau d\tau \right) ; \ \ t \geq 0,
\end{equation}
where $\mathcal{T}$ is the time-ordering operator \cite{BreuerPetr}. Now, the dynamics is much more complicated, and the conditions for $\mathcal{L}(t)$ that guarantee complete positivity of $\Lambda(t)$ for all $t \geq 0$ are not known. A sufficient condition is the positivity of all decoherence rates
$\gamma_\alpha(t)\geq 0$ for all $t\geq 0$. This requirement is very strong and implies not only that $\Lambda(t)$ is CPTP but that all propagators
\begin{equation}\label{}
V(t,s) = \mathcal{T} \exp\left( \int_s^t \mathcal{L}(\tau) d\tau \right) ; \ \ t \geq s,
\end{equation}
are CPTP, as well. Such evolution is called {\it CP-divisible}, and it satisfies the equation $\Lambda(t) = V(t,s)\Lambda(s)$. CP-divisibility of $\Lambda(t)$ is often considered to be a definition of Markovianity \cite{RHP}. If the evolution is commutative, i.e. $\mathcal{L}(t)\mathcal{L}(\tau) = \mathcal{L}(\tau)\mathcal{L}(t)$ for any pair $t$ and $\tau$, then we can drop the chronological operator $\mathcal{T}$ from eq. (\ref{T-exp}). In this case, a sufficient condition for $\mathcal{L}(t)$ to generate a CP-divisible dynamical map states that $\int_0^t \mathcal{L}(\tau)d\tau$ is a time-dependent GKSL generator for any $t>0$.

A standard example in open quantum dynamics is a quantum evolution with only one decoherence channel. Consider the well-known qubit decoherence governed by
\begin{equation}\label{}
  \mathcal{L}(t)[\rho] = \frac 12 \gamma(t)(\sigma_3 \rho \sigma_3 - \rho) .
\end{equation}
In this case, the necessary and sufficient condition for legitimacy of the solution states that $\Gamma(t) = \int_0^t \gamma(\tau)d\tau \geq 0$ for all $t> 0$. This condition is no longer necessary if there are more decoherence channels. An instructive example was provided in \cite{ENM}, where the authors considered the Pauli channel $\Lambda(t)$ generated by
\begin{equation}\label{123}
\mathcal{L}(t)[\rho]=\frac 12 \sum_{\alpha=1}^3\gamma_\alpha(t)(\sigma_\alpha \rho\sigma_\alpha-\rho)
\end{equation}
with two positive $\gamma_1(t)=\gamma_2(t)=1$ and one always negative $\gamma_3(t)=-\tanh t$ decoherence rates. Since $\gamma_3(t) < 0$, its integral $\Gamma_3(t) < 0$. Nevertheless, as shown in \cite{ENM}, the corresponding map is CPTP. Since one of the rates is permanently negative, the authors called this evolution {\it eternally non-Markovian}. Surprisingly, this evolution, which seems to be highly non-Markovian, has the monotonicity property
\begin{equation}\label{BLP}
  \frac{d}{dt} \| \Lambda(t)(\rho_1 - \rho_2)\|_1 \leq 0
\end{equation}
for any pair of qubit density operators $\rho_1,\rho_2$ ($\|X\|_1 = {\rm Tr}\sqrt{XX^\dagger}$ denotes the trace norm of $X$). This property is considered an alternative concept of Markovianity \cite{BLP}. Moreover, Megier et al. \cite{Nina} have shown that the Pauli channel $\Lambda(t)$ generated by $\mathcal{L}(t)$ from eq. (\ref{123}) arises from a simple mixture of two Markovian semigroups,
\begin{equation}\label{ENM_2}
\Lambda(t)=\frac 12 \left(e^{2t\mathcal{L}_1}+e^{2t\mathcal{L}_2}\right),\quad \mathcal{L}_\alpha[\rho]=\frac 12 (\sigma_\alpha \rho\sigma_\alpha-\rho).
\end{equation}

In this paper, we generalize the eternally non-Markovian evolution of a qubit. First, we consider the mixture of legitimate qubit dynamical maps generated by time-local generators. Next, this evolution is generalized even further to the qudit dynamics described by the generalized Pauli channels \cite{Ruskai}. As a special case, we analyze the convex combination of Markovian semigroups. We show how to realize this evolution as a classical Markov process. For dimension $d=3$, we explicitly give the range of parameters that lead to a CP or P-divisible dynamical map. Moreover, we construct an example of a quantum evolution of a $d$-level system with a large number of permanently negative decoherence rates. Finally, we find that all the rates cam be temporarily negative and the evolution is still given by a legitimate dynamical map.

\section{Pauli channels}

Consider the qubit evolution
\begin{equation}\label{PC}
\Lambda(t)[\rho] = \sum_{\alpha=0}^{3} p_\alpha(t) \sigma_\alpha \rho \sigma_\alpha
\end{equation}
generated by the time-local generator from eq. (\ref{123}). In terms of the eigenvalues $\lambda_k(t)$ of $\Lambda(t)$ to the eigenvectors $\sigma_k$, the probability 4-vector $p_\alpha(t)$ reads \cite{Filip}
\begin{equation}\label{}
  p_0(t) = \frac 14 (1 + \lambda_1(t) + \lambda_2(t) + \lambda_3(t))
\end{equation}
and
\begin{equation}\label{}
  p_k(t) = \frac 14 (1 + 2\lambda_k(t) - \lambda_1(t) - \lambda_2(t) - \lambda_3(t)).
\end{equation}
Finally, one has
\begin{equation}\label{}
  \lambda_k(t) = e^{\Gamma_k(t) - \Gamma(t)} ,
\end{equation}
where $\Gamma(t) = \sum_{k=1}^3 \Gamma_k(t)$. 

Let us analyze the following mixture of legitimate qubit dynamics,
\begin{equation}\label{xxx}
\Lambda(t) = x_1 e^{w_1(t) \mathcal{L}_1}  + x_2 e^{w_2(t) \mathcal{L}_2} +  x_3 e^{w_3(t) \mathcal{L}_3}  ,
\end{equation}
where $w_k(t) \geq 0$ for all $t\geq 0$, and $(x_1,x_2,x_3)$ is a probability vector. This model has been studies in \cite{Nina} in the special
case where $w_1(t)=w_2(t)=w_3(t)=2t$. One finds
\begin{equation}\label{}
  \left( \begin{array}{c}
    \gamma_1 \\
    \gamma_2 \\
    \gamma_3
  \end{array} \right) = \frac 12 \left( \begin{array}{ccc}
-1 & 1 & 1 \\
1 & -1 & 1 \\
1 & 1 & -1
\end{array} \right) \left( \begin{array}{c}
    \mu_1 \\
    \mu_2 \\
    \mu_3
  \end{array} \right) ,
\end{equation}
where
\begin{eqnarray}
\mu_1 &=& \frac{ x_2 \dot{w}_2 e^{-w_2} + x_3 \dot{w}_3 e^{-w_3} }{ x_1 +  x_2  e^{-w_2} + x_3  e^{-w_3} } , \nonumber \\
\mu_2 &=& \frac{ x_1 \dot{w}_1 e^{-w_1} + x_3 \dot{w}_3 e^{-w_3} }{ x_2 +  x_1  e^{-w_1} + x_3  e^{-w_3} } , \\
\mu_3 &=& \frac{ x_1 \dot{w}_1 e^{-w_1} + x_2 \dot{w}_2 e^{-w_2} }{ x_3 +  x_1  e^{-w_1} + x_2  e^{-w_2} } . \nonumber
\end{eqnarray}
The dynamical map given by eq. (\ref{xxx}) is P-divisible if and only if
\begin{equation}
\mu_k(t) \geq 0
  \end{equation}
for $k=1,2,3$. It is, therefore, clear that if all $\dot{w}_k(t) \geq 0$, then $\Lambda(t)$ is P-divisible. However, if $\dot{w}_k(t) \ngeq 0$, then it may happen that temporarily all rates $\gamma_k(t)$ are strictly negative. Note that for $w_1(t)=w_2(t)=w_3(t)\equiv w(t)$, $\Lambda(t)$ is P-divisible if and only if $\dot{w}(t) \geq 0 $ for all $t\geq 0$. Interestingly, if $\dot{w}(t) \geq 0 $, then $ \gamma_1(t)=\gamma_2(t) \geq 0$ and $ \gamma_3(t) \leq 0$. However, if $\dot{w}(t) \leq 0 $, then $\gamma_1(t)=\gamma_2(t) \leq 0$ and $ \gamma_3(t) \geq 0$. One easily checks that the corresponding dynamical map
\begin{equation}\label{*}
\Lambda(t) = \frac 12 \left( e^{w(t) \mathcal{L}_1}  + e^{w(t) \mathcal{L}_2} \right)
\end{equation}
recovers the eternaly non-Markovian evolution from \cite{ENM} for $w(t)=2t$.

Now, the map from eq. (\ref{xxx}) corresponds to the Pauli channel with
\begin{equation}\label{}
  p_0(t) = \frac 12 \left( 1 + x_1 e^{-w_1(t)} + x_2 e^{-w_2(t)} + x_3 e^{-w_3(t)} \right)
\end{equation}
and
\begin{equation}\label{}
  p_k(t) = \frac{x_k}{2} \left(1- e^{-w_k(t)} \right) , \ \ k=1,2,3. 
\end{equation}
Interestingly, if all $\dot{w}_k(t) \geq 0$, then $p_\alpha(t)$ satisfy the classical equation
\begin{equation}\label{clas}
\frac{d}{dt} \left( \begin{array}{c}
p_0 \\
p_1 \\
p_2 \\
p_3
\end{array}\right) = \frac 12 \mathfrak{L}(t) \left( \begin{array}{c}
p_0 \\
p_1 \\
p_2 \\
p_3
\end{array}\right)
\end{equation}
with the Markovian generator
\begin{equation}\label{}
\mathfrak{L}=\left( \begin{array}{cccc}
-W & 2\dot{w}_1-W & 2\dot{w}_2-W & 2\dot{w}_3-W \\
\dot{w}_1x_1 & -(2-x_1)\dot{w}_1 & \dot{w}_1x_1 & \dot{w}_1x_1 \\
\dot{w}_2x_2 & \dot{w}_2x_2 & -(2-x_2)\dot{w}_2 & \dot{w}_2x_2 \\
\dot{w}_3x_3 & \dot{w}_3x_3 & \dot{w}_3x_3 & -(2-x_3)\dot{w}_3
\end{array} \right),
\end{equation}
where $W(t):=\dot{w}_1(t)x_1+\dot{w}_2(t)x_2+\dot{w}_3(t)x_3$. Note that eq. (\ref{clas}) has a form of a classical Pauli master equation with time-dependent rates. This equation is legitimate provided that all $\dot{w}_\alpha(t)\geq 0$ and $W(t)\leq 2\min_\alpha\dot{w}_\alpha(t)$. Interestingly, as long as $w_1(t)=w_2(t)=w_3(t)=w(t)$, many of the rates vanish, and the classical master equation reduces to
\begin{equation}
\frac{d}{dt} \left( \begin{array}{c}
p_0 \\
p_1 \\
p_2 \\
p_3
\end{array}\right) = \frac{\dot{w}}{2}
\left( \begin{array}{cccc}
-1 & 1 & 1 & 1 \\
x_1 & -1 & 0 & 0 \\
x_2 & 0 & -1 & 0 \\
x_3 & 0 & 0 & -1
\end{array} \right)
\left( \begin{array}{c}
p_0 \\
p_1 \\
p_2 \\
p_3
\end{array}\right).
\end{equation}
This form is very similar to the result of \cite{Nina}.

\section{Generalized Pauli channels}

Consider the evolution of a qudit in a $d$-dimensional Hilbert space $\mathcal{H}$ that admits the maximal number $N(d)=d+1$ of mutually unbiased bases (MUBs) $\{\psi_0^{(\alpha)},\dots,\psi_{d-1}^{(\alpha)}\}$ \cite{Wootters,MAX}. Recall that two bases are mutually unbiased if
\begin{equation}
\big\<\psi_k^{(\alpha)}\big|\psi_l^{(\alpha)}\big\>=\delta_{kl},\quad
\big|\big\<\psi_k^{(\alpha)}\big|\psi_l^{(\beta)}\big\>\big|^2=
\frac 1d,\quad\alpha\neq\beta.
\end{equation}
Now, assume that the generator of the evolution has the form
\begin{equation}\label{GEN2}
\mathcal{L}(t)=\sum_{\alpha=1}^{d+1}\gamma_\alpha(t)\mathcal{L}_\alpha,\qquad\mathcal{L}_\alpha=\Phi_\alpha-\oper,
\end{equation}
where
\begin{equation}
\Phi_\alpha[X]=\sum_{k=0}^{d-1}P_k^{(\alpha)}XP_k^{(\alpha)},
\end{equation}
and $P_k^{(\alpha)}:=|\psi_k^{(\alpha)}\>\<\psi_k^{(\alpha)}|$ denote the rank-1 projectors onto the MUB vectors. The solution of the time-local master equation $\dot{\Lambda}(t)=\mathcal{L}(t)\Lambda(t)$ with generator (\ref{GEN2}) is the generalized Pauli channel \cite{Ruskai,mub_final}
\begin{equation}\label{GPC}
\Lambda(t)=\frac{dp_0(t)-1}{d-1}\oper+\frac{d}{d-1}\sum_{\alpha=1}^{d+1}p_\alpha(t)\Phi_\alpha,
\end{equation}
where $p_\alpha(t)$ denotes the probability $(d+2)$-vector with the components
\begin{align}
p_0(t)&=\frac{1}{d^2}\left[1+(d-1)\sum_{\alpha=1}^{d+1}\lambda_\alpha(t)\right],\label{c1}\\
p_\alpha(t)&=\frac{d-1}{d^2}\left[1+d\lambda_\alpha(t)-\sum_{\beta=1}^{d+1} \lambda_\beta(t)\right].\label{c2}
\end{align}
The eigenvalues of $\Lambda(t)$ to the eigenvectors
\begin{equation}\label{U}
U_{\alpha}^k=\sum_{l=0}^{d-1}\omega^{kl}P_l^{(\alpha)},\quad k=1,\ldots,d-1,
\end{equation}
with $\omega = e^{2\pi i/d}$ are given by
\begin{equation}
  \lambda_\alpha(t) = \exp[\Gamma_\alpha(t) - \Gamma(t)] ,
\end{equation}
where $\Gamma(t) = \sum_{k=1}^{d+1} \Gamma_k(t)$. For $d=2$, one reproduces the Pauli channel from eq. (\ref{PC}).

The qubit map from eq. (\ref{xxx}) can be straightforwardly generalized to
\begin{equation}\label{yyy}
\Lambda(t) = \sum_{\alpha=1}^{d+1}x_\alpha e^{w_\alpha(t) \mathcal{L}_\alpha}
\end{equation}
with $w_\alpha(t)\geq 0$ for all $t \geq 0$ and $w_\alpha(0)=0$. The associated decoherence rates are given by
\begin{equation}\label{gammas2}
\gamma_\alpha(t)=
\frac 1d \sum_{\beta=1}^{d+1}\mu_\beta(t)-\mu_\alpha(t),
\end{equation}
where
\begin{equation}
\mu_\alpha(t)=\frac{\sum_{\beta=1}^{d+1}x_\beta\dot{w}_\beta e^{-w_\beta(t)}-x_\alpha\dot{w}_\alpha(t)e^{-w_\alpha(t)}}
{x_\alpha(1-e^{-w_\alpha(t)})+\sum_{\beta=1}^{d+1}x_\beta e^{-w_\beta(t)}}.
\end{equation}
Unluckily, the necessary and sufficient conditions for P-divisibility of the generalized Pauli channels are not known. However, we know that the dynamical map from eq. (\ref{yyy}) is not P-divisible if
\begin{equation}\label{}
\mu_\alpha(t)\ngeq 0
\end{equation}
for any $\alpha=1,\ldots,d+1$. Clearly, if all $\dot{w}_\alpha(t)\leq 0$ at some time $t$, then $\Lambda(t)$ is not P-divisible.

The dynamical map in eq. (\ref{yyy}) corresponds to the generalized Pauli channel with
\begin{equation}\label{}
  p_0(t) = \frac 1d \left( 1 + [d-1]\sum_{\alpha=1}^{d+1}x_\alpha e^{-w_\alpha(t)}\right)
\end{equation}
and
\begin{equation}\label{}
  p_k(t) = \frac{d-1}{d}x_k \left(1- e^{-w_k(t)} \right) , \ \ k=1,\ldots,d+1. 
\end{equation}
Interestingly, if all $\dot{w}_k(t)\geq 0$, then $p_\alpha(t)$ satisfy the classical equation
\begin{equation}\label{}
\dot{p}_\alpha=\frac 1d \sum_{\beta=1}^{d+1}\mathfrak{L}_{\alpha\beta}p_\beta
\end{equation}
with the Markovian generator
\begin{equation}\label{}
\begin{split}
&\mathfrak{L}_{00}(t)=-W(t),\\
&\mathfrak{L}_{0k}(t)=-W(t)+d\dot{w}_\beta(t),\\
&\mathfrak{L}_{k0}(t)=\mathfrak{L}_{kl}(t)=(d-1)x_k\dot{w}_k(t),\\
&\mathfrak{L}_{kk}(t)=-[d(1-x_k)+x_k]\dot{w}_k(t),
\end{split}
\end{equation}
where $W(t):=(d-1)\sum_{\alpha=1}^{d+1}\dot{w}_\alpha(t)x_\alpha$. For example, if $d=3$ and all $w_k(t)=w(t)$, then
\begin{equation}
\mathfrak{L} = \frac{\dot{w}}{3}
\left( \begin{array}{ccccc}
-2 & 1 & 1 & 1 & 1 \\
x_1 & -1 & 0 & 0 & 0 \\
x_2 & 0 & -1 & 0 & 0 \\
x_3 & 0 & 0 & -1 & 0 \\
x_4 & 0 & 0 & 0 & -1
\end{array} \right).
\end{equation}

Actually, one can have all $\gamma_\alpha(t)\leq 0$ but not eternally. Indeed, if $w_\alpha(t)=w(t)$ and $x_\alpha=\frac{1}{d+1}$, then the corresponding
\begin{equation}
\gamma_\alpha(t)=\frac{\dot{w}(t)}{d+e^{w(t)}}.
\end{equation}
Assume that, for some $t_\ast>0$, one has $\dot{w}(t_\ast)<0$ even though $w(t_\ast)>0$. Then, the evolution is non-Markovian with all $d+1$ decoherence rates that are negative for $t=t_\ast$, while the dynamical map $\Lambda(t)$ is still legitimate (CPTP). It is noteworthy that the rates do not stay negative forever because $w(t)\geq 0$ for all $t\geq 0$.

\section{Convex combination of Markovian semigroups}

As a special case, let us analyze the mixture of dynamical semigroups. Interestingly, even a simple mixture may lead to a much more involved quantum evolution than the dynamics of its components. Let us take the generalized Pauli channel that is a convex combination of Markovian semigroups,
\begin{equation}\label{CCGPC}
\begin{split}
\Lambda(t)&=\sum_{\alpha=1}^{d+1}x_\alpha e^{rt\mathcal{L}_\alpha}\\&=
e^{-rt}\oper+(1-e^{-rt})\sum_{\alpha=1}^{d+1}x_\alpha \Phi_\alpha,
\end{split}
\end{equation}
whose corresponding probability distribution is given by
\begin{equation}\label{prob}
\begin{split}
p_0(t)&=\frac 1d (1+[d-1]e^{-rt}),\\
p_\alpha(t)&=\frac{d-1}{d}(1-e^{rt})x_\alpha.
\end{split}
\end{equation}
Its eigenvalues
\begin{equation}
\lambda_\alpha(t)=e^{-rt}+\left(1-e^{-rt}\right)x_\alpha
\end{equation}
have a relatively simple form and each depend on a single $x_\alpha$.
Yet, the resulting map is no longer a semigroup. Instead, it solves the master equation $\dot{\Lambda}(t)=\mathcal{L}(t)\Lambda(t)$, where the time-local generator has the decoherence rates 
\begin{equation}\label{gammas}
\gamma_\alpha(t)=-\frac{(1-x_\alpha)r} {1+\left(e^{rt}-1\right)x_\alpha}+\frac{r}{d}
\sum_{\beta=1}^{d+1}\frac{1-x_\beta} {1+\left(e^{rt}-1\right)x_\beta} 
\end{equation}
that no longer have to be positive. One has
\begin{equation*}
\begin{bmatrix}
    \gamma_1 \\
    \gamma_2 \\
    \vdots \\
    \gamma_{d+1}
  \end{bmatrix} \!=\! \frac rd \begin{bmatrix}
-(d-1) & 1 & \ldots & 1 \\
1 & -(d-1) & \ldots & 1 \\
\vdots & \vdots & \ddots & \vdots \\
1 & 1 & \ldots & -(d-1)
\end{bmatrix}\!
\begin{bmatrix}
    \mu_1 \\
    \mu_2 \\
    \vdots \\
    \mu_{d+1}
  \end{bmatrix},
\end{equation*}
where
\begin{equation}\label{mu}
\mu_\alpha(t)=\frac{1-x_\alpha} {1+\left(e^{rt}-1\right)x_\alpha}.
\end{equation}
If $\gamma_\alpha(t)\geq 0$ for all $t\geq 0$, then the resulting evolution is CP-divisible. Sufficient conditions for P-divisibility of the generalized Pauli dynamical maps have been derived in \cite{ICQC}. They read as follows,
\begin{equation}\label{suf}
[d-2(k-1)]\gamma_\beta(t)+[d+2(k-1)]\gamma_\alpha(t)\geq 0,
\end{equation}
where $\alpha=1,\ldots,k\leq\frac{d+1}{2}$ and $\beta=k+1,\ldots,d+1$ number negative and positive decoherence rates, respectively.

\begin{Example}\label{Ex8}
For the maximally mixed probability vector $x_\alpha=\frac{1}{d+1}$, the time-local generator $\mathcal{L}(t)$ is determined by
\begin{equation}
\gamma_\alpha(t)=\frac{r}{d+e^{rt}}\geq 0,
\end{equation}
which always correspond to the Markovian evolution.
\end{Example}

%\begin{Remark}
%In principle, another possible mixture of Markovian semigroups is
%\begin{equation}
%\Lambda(t)=\sum_{\alpha=1}^{d+1}x_\alpha e^{\eta_\alpha t\mathcal{L}_\beta}
%=\frac 1d \Bigg[(1+[d-1]\sum_{\alpha=1}^{d+1}x_\alpha e^{-\eta_\alpha t})\oper+(1-\sum_{\alpha=1}^{d+1}x_\alpha e^{-\eta_\alpha t})\mathbb{U}_\beta\Bigg].
%\end{equation}
%\end{Remark}

\begin{Example}\label{ENM_ex}
If $x_\alpha=1/d$ for $\alpha=1,\dots,d$ and $x_{d+1}=0$, the dynamical map is generated by $\mathcal{L}(t)$ with
\begin{equation}\label{EM_gammas}
\begin{split}
\gamma_\alpha(t)&=\frac{r}{d},\quad \alpha=1,\dots,d,\\
\gamma_{d+1}(t)&=-\frac{r}{d}\frac{(d-1)\left(e^{rt}-1\right)}{e^{rt}-1+d}
\leq 0,
\end{split}
\end{equation}
where one decoherence rate is eternally negative.
\end{Example}

\begin{Example}\label{EX_x}
For the choice $x_1=x_2=1/2$ and $x_\alpha=0$ when $\alpha=3,\ldots,d+1$, the associated $\mathcal{L}(t)$ has
\begin{equation}
\begin{split}
&\gamma_1(t)=\gamma_2(t)=\frac rd \frac{d-1+e^{-rt}}{1+e^{-rt}}\geq 0,\\
&\gamma_\alpha(t)=-\frac rd \tanh\frac{rt}{2}\leq 0,\qquad\alpha=3,\dots,d+1.
\end{split}
\end{equation}
Hence, there exists a legitimate time-local generator with $d-1$ identical eternally negative $\gamma_\alpha(t)$.
\end{Example}

Actually, there is an entire family of channels describing the eternally non-Markovian evolution with $1\leq k\leq d-1$ negative decoherence rates.

\begin{Example}\label{kdr}
Let us take $x_\alpha=0$ for $\alpha=1,\ldots,k$ and $x_\alpha=\frac{1}{d+1-k}$ whenever $\alpha=k+1,\ldots,d+1$. The corresponding decoherence rates are given by
\begin{equation}
\begin{split}
&\gamma_\alpha(t)=\frac{r}{d}\frac{d+k(e^{rt}-1)}{d-k+e^{rt}}\geq 0,\qquad \alpha=k+1,\ldots,d+1,\\
&\gamma_\alpha(t)=-\frac{r}{d}\frac{(d-k)(e^{rt}-1)}{d-k+e^{rt}}\leq 0,\qquad \alpha=1,\ldots,k.
\end{split}
\end{equation}
Hence, the resulting evolution has $k$ eternally negative rates, where $1\leq k\leq d-1$.
\end{Example}

The above examples have already been considered in the special cases of $r=d$ \cite{memory}, $r=2,d$ \cite{mub_final}, and $r=d+1-k$ \cite{ICQC}, respectively.

By manipulating the value of $r$, one can change the magnitude but not the sign of the rates. Moreover, the $r$-dependence is highly non-trivial. With the increase of the dimension $d$, one observes an interesting asymptotic behavior. In Example \ref{ENM_ex}, if $r\neq d$, then there are $d$ positive decoherence rates that are very small and one rate that is slightly negative. For $d=r$, however, all $d$ positive rates are equal to identity, whereas the negative rate approaches $-\infty$. The situation is quite different in Example \ref{EX_x}. There, if $r\neq d$, one has two decoherence rates equal to unity and $d-1$ rates that are slightly negative. Meanwhile, $r=d$ leads to two positive rates that approach $\infty$ and $d-1$ rates that are equal to $-1$. Finally, in Example \ref{kdr}, for $r\neq d$, there are $k$ very small positive and $d+1-k$ very large negative rates. If $r=d$, we have $\gamma_\alpha(t)\to k$ for $\alpha>k$ and $\gamma_\alpha(t)\to-\infty$ for $\alpha\leq k$. The analysis of Example \ref{kdr} is based on the assumption that $k$ does not depend on $d$.

\section{Classical Markov process}

Even though a mixture of Markovian dynamical maps is sometimes generated with negative decoherence rates, the evolution can be simulated with a classical Markov process \cite{Nina}. Indeed, $p_\alpha$ given in eq. (\ref{prob}) solve the rate equations
\begin{equation}
\begin{split}
&\dot{p}_0=-(d-1)p_0+\sum_{\beta=1}^{d+1}p_\beta,\\
&\dot{p}_\alpha=-p_\alpha+(d-1)p_0x_\alpha.
\end{split}
\end{equation}
Note that the above rate equations have the form of the classical Pauli master equation \cite{vanKampen}
\begin{equation}
\dot{p}_\alpha(t)=\sum_\beta\Big[\Gamma_{\beta\to\alpha}p_\alpha(t)
-\Gamma_{\alpha\to\beta}p_\beta(t)\Big]
\end{equation}
with the only non-vanishing rates $\Gamma_{0\to \alpha}=(d-1)x_\alpha$, $\Gamma_{\alpha\to 0}=1$, $\alpha=1,\ldots,d+1$, being positive and time-independent. For $d=3$, the evolution equation can be also rewritten in the matrix form
\begin{equation}
\frac{\der}{\der t}\begin{pmatrix}
p_0(t)\\ p_1(t) \\ p_2(t) \\ p_3(t) \\ p_4(t)
\end{pmatrix}
=
\begin{pmatrix}
-2 & 1 & 1 & 1 & 1 \\
2x_1 & -1 & 0 & 0 & 0 \\
2x_2 & 0 & -1 & 0 & 0 \\
2x_3 & 0 & 0 & -1 & 0 \\
2x_4 & 0 & 0 & 0 & -1 
\end{pmatrix}
\begin{pmatrix}
p_0(t)\\ p_1(t) \\ p_2(t) \\ p_3(t) \\ p_4(t)
\end{pmatrix}.
\end{equation}

The probability distribution $p_\alpha$ from eq. (\ref{prob}) is also the solution of the following rate equations,
\begin{equation}
\dot{p}_0(t)=\frac 1d 
\left[-(d-1)\gamma_0(t)p_0(t)+\sum_{\beta=1}^{d+1}\gamma_\beta(t)p_\beta(t)\right],
\end{equation}
\begin{equation}
\begin{split}
\dot{p}_\alpha(t)=\frac 1d \Bigg\{\Big[(d-2)\gamma_\alpha(t) -(d-1)\gamma_0(t)\Big]p_\alpha(t)&\\
+\sum_{\beta\neq\alpha}p_\beta(t)\Big[\gamma_0(t)-\gamma_\alpha(t)-\gamma_\beta(t)\Big]
&\\\qquad\qquad\qquad+(d-1)\gamma_\alpha(t)p_0(t)&\Bigg\}.
\end{split}
\end{equation}
For $d=3$, this takes a relatively simple form,
\begin{equation}
\frac{\der}{\der t}\begin{pmatrix}
p_0(t)\\ p_1(t) \\ p_2(t) \\ p_3(t) \\ p_4(t)
\end{pmatrix}
=\frac 13
\mathcal{A}(t)
\begin{pmatrix}
p_0(t)\\ p_1(t) \\ p_2(t) \\ p_3(t) \\ p_4(t)
\end{pmatrix},
\end{equation}
with $\mathcal{A}(t)$ equal to
\begin{equation}
\begin{pmatrix}
-2\gamma_0 & \gamma_1 & \gamma_2 & \gamma_3 & \gamma_4 \\
2\gamma_1 & \gamma_1-2\gamma_0 & \gamma_3+\gamma_4 & \gamma_2+\gamma_4 & \gamma_2+\gamma_3 \\
2\gamma_2 & \gamma_3+\gamma_4 & \gamma_2-2\gamma_0 & \gamma_1+\gamma_4 & \gamma_1+\gamma_3 \\
2\gamma_3 & \gamma_2+\gamma_4 & \gamma_1+\gamma_4 & \gamma_3-2\gamma_0 & \gamma_1+\gamma_2 \\
2\gamma_4 & \gamma_2+\gamma_3 & \gamma_1+\gamma_3 & \gamma_1+\gamma_2 & \gamma_4-2\gamma_0 
\end{pmatrix}
\end{equation}
(time-dependence omitted for the sake of clarity).
This time, however, the associated classical Pauli master equation
\begin{equation}
\dot{p}_\alpha(t)=\frac 1d \sum_\beta\Big[\gamma_{\beta\to\alpha}(t)p_\alpha(t)
-\gamma_{\alpha\to\beta}(t)p_\beta(t)\Big]
\end{equation}
has time-dependent rates 
\begin{equation}
\begin{split}
&\gamma_{0\to\alpha}(t)=\gamma_\alpha(t),\\
&\gamma_{\alpha\to 0}(t)=(d-1)\gamma_\alpha(t),\\
&\gamma_{\alpha\to\beta}(t)=\gamma_0(t)-\gamma_\alpha(t)-\gamma_\beta(t)
\end{split}
\end{equation}
that can become negative. Hence, it does not describe a legitimate Markov process.

\section{Markovianity and non-Markovianity regions}

\begin{figure}[ht!]
  \centering
\tiny
       \includegraphics[width=0.5\textwidth]{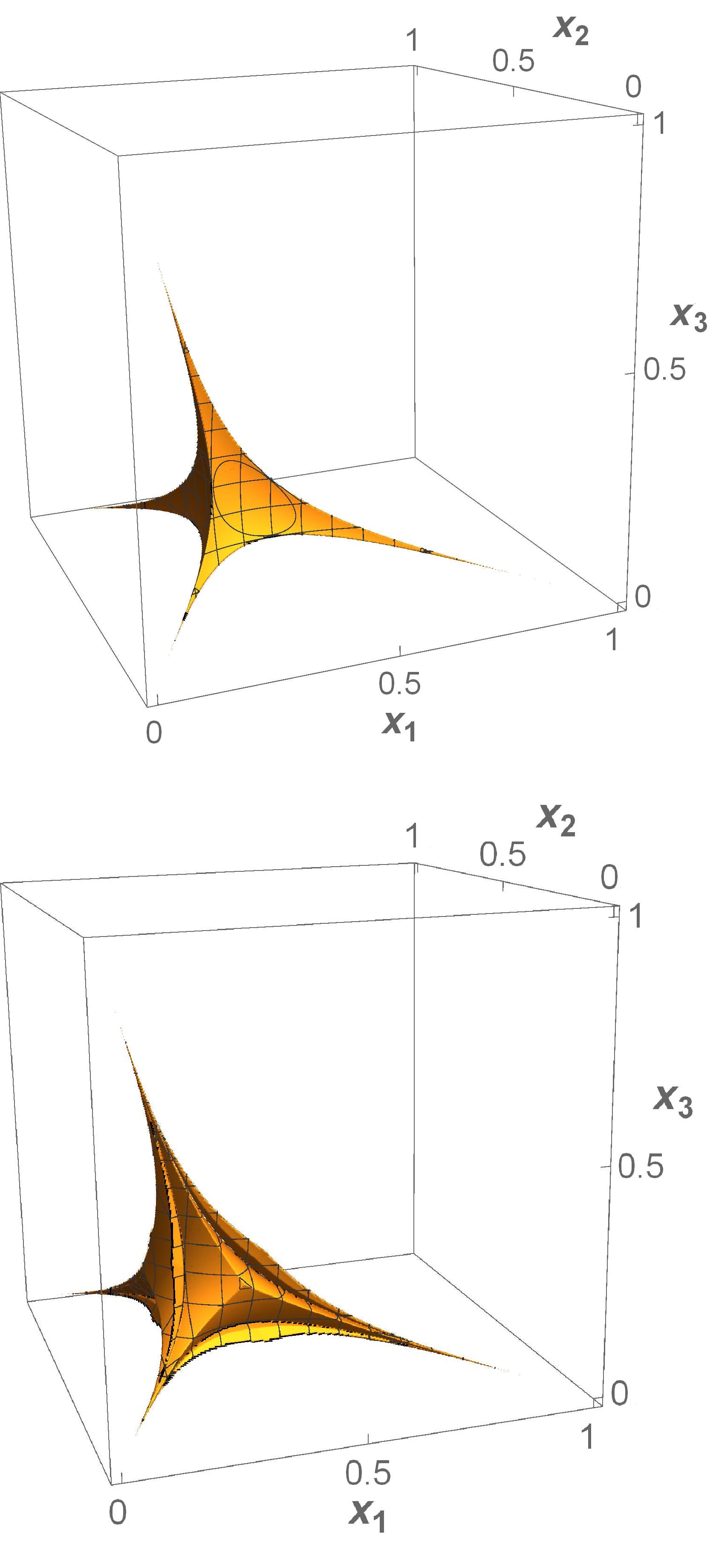}
\caption{\it\small The range of parameters $x_1,\ x_2,\ x_3$ in $d=3$ for which $\gamma_\alpha(t)\geq 0$ at all times $t\geq 0$ (up) and for which the necessary P-divisibility conditions hold (down).}\label{CC_CP}
\end{figure}

Even though $\Lambda(t)$ given by eq. (\ref{CCGPC}) is a convex combination of CP-divisible dynamical maps, the resulting channel is CP-divisible only for certain values of $x_\alpha$. For $d=2$, every convex combination of Markovian semigroups is P-divisible. It is still unknown whether this property carries over to $d\geq 3$. However, $\Lambda(t)$ always satisfies the necessary P-divisibility conditions \cite{mub_final}
\begin{equation}
\frac{1}{r}\sum_{\beta\neq\alpha}\gamma_\beta(t)=\mu_\alpha(t)
\geq 0.
\end{equation}
The sufficient conditions from ineq. (\ref{suf}) reduce to
\begin{equation}\label{suf2}
2\sum_{\nu=1}^{d+1}\mu_\nu(t)\geq
[d+2(k-1)]\mu_\alpha(t)+[d-2(k-1)]\mu_\beta(t),
\end{equation}
where again $\alpha=1,\ldots,k\leq\frac{d+1}{2}$ and $\beta=k+1,\ldots,d+1$ number negative and positive decoherence rates, respectively. From among Examples 1--4, none has the decoherence rates that satisfy the sufficient conditions for P-divisibility.

The CP-divisibility region can be derived analytically. The necessary and sufficient conditions for CP-divisibility are $\gamma_\alpha(t)\geq 0$, which can be rewritten into
\begin{equation}\label{dd}
\left(\prod_{\nu=1}^{d+1}x_\nu\right)\left[\sum_{\beta=1}^{d+1}\frac{1}{x_\beta}-\frac{d}{x_\alpha} -1\right]\geq 0
\end{equation}
after multiplying (\ref{gammas}) by $e^{-drt}\prod_{\mu=1}^{d+1}\left[1+\left(e^{rt}-1\right)x_\mu\right]$ and taking the limit $t\to\infty$. The border of the CP-divisibility region corresponds to the equality. The hyperplanes $x_\alpha=x_\beta$ give $d+1$ symmetry planes, which cross at $x_\alpha=1/d$. Analogically, from the sufficient conditions for P-divisibility in eq. (\ref{suf2}), one obtains the following P-divisibility region,
\begin{equation}\label{ee}
\begin{split}
\left(\prod_{\nu=1}^{d+1}x_\nu\right)\Bigg[-\frac{d+2(k-1)}{2x_\alpha}
&-\frac{d-2(k-1)}{2x_\beta}\\&\quad+\sum_{\mu=1}^{d+1}\frac{1}{x_\mu}-1\Bigg]
\geq 0.
\end{split}
\end{equation}
However, it is not straightforward to find the region border, as the left hand-side of (\ref{ee}) depends on $k$.

\begin{Example}\label{reg}
For $d=3$, the border of CP-divisibility region (\ref{dd}) is given by
\begin{equation}
x_4=\begin{cases}
&\displaystyle\frac{x_1x_2x_3}{-x_1x_2+x_1x_3-x_2x_3+2x_1x_2x_3},\\[12pt] &0\leq x_2\leq x_3\leq 1,\quad\frac{x_2x_3}{-x_2+x_3+x_2x_3}\leq x_1\leq 1,\\[12pt]
&\displaystyle\frac{x_1x_2x_3}{-x_1x_2-x_1x_3+x_2x_3+2x_1x_2x_3},\\[12pt] & 0\leq x_2,x_3\leq 1,\quad 0\leq x_1\leq\frac{x_2x_3}{x_2+x_3-x_2x_3},\\[12pt]
&\displaystyle\frac{x_1x_2x_3}{x_1x_2-x_1x_3-x_2x_3+2x_1x_2x_3},\\[12pt] & 0\leq x_3\leq x_2\leq 1,\quad\frac{x_2x_3}{x_2-x_3+x_2x_3}\leq x_1\leq 1,\\[12pt]
&\displaystyle\frac{x_1x_2x_3}{x_1x_2+x_1x_3+x_2x_3-2x_1x_2x_3},\\[12pt] & 0\leq x_1,x_2,x_3\leq 1.
\end{cases}
\end{equation}
\end{Example}

For $d=r=3$, both CP and P-divisibility regions can be plotted. In Fig. \ref{CC_CP}, we see that neither of them represents a convex set. Moreover, the regions are symmetric with respect to the change of indices of $x_\alpha$. Additionally, the P-divisibility region possesses some discontinuities that are absent in the case where $d=r=2$ \cite{Nina}. This could indicate that the sufficient conditions for P-divisibility in (\ref{suf}) are not necessary and sufficient for $d=3$.

\begin{figure}[ht!]
  \centering
\tiny
       \includegraphics[width=0.5\textwidth]{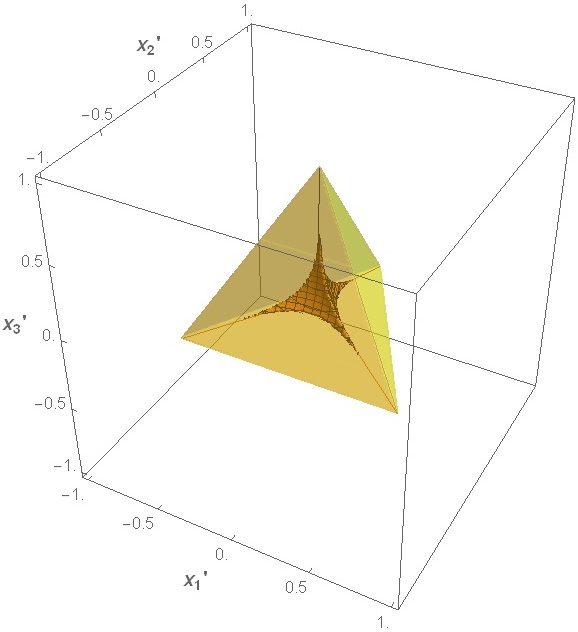}
\caption{\it\small The CP-divisibility region for $d=3$ plotted on a simplex.}
\label{simplex}
\end{figure}

In Fig. \ref{simplex}, we show the CP-divisibility region on a simplex.
The simplex coordinates read
\begin{align}
&x_\alpha^\prime=\frac{\alpha x_{\alpha+1}-\sum_{\beta=1}^\alpha x_\beta}{\prod_{\mu=1}^\alpha\sqrt{1+\mu}},\\
&x_{d+1}^\prime=\frac{1}{\sqrt{d+1}},
\end{align}
which for $d=3$ reduces to
\begin{align}
&x_1^\prime=\frac{-x_1+x_2}{\sqrt{2}},\\
&x_2^\prime=\frac{-x_1-x_2+2x_3}{\sqrt{6}},\\
&x_3^\prime=\frac{-x_1-x_2-x_3+3x_4}{2\sqrt{3}},\\
&x_4^\prime=\frac 12.
\end{align}
We see that the Markovian evolution constitutes a relatively small portion compared to the set of admissible quantum evolutions. Moreover, the region of CP-divisible dynamical maps has the same symmetry planes as the simplex.

\FloatBarrier

\section{Conclusions}

We generalized the eternally non-Markovian evolution of a qubit to the $d$-level systems. In particular, we showed how to construct the generalized Pauli channels by taking the mixture of legitimate dymanical maps. As a special case, we analyzed the convex combination of Markovian semigroups. We also provided several interesting examples of non-Markovian and eternally non-Markovian evolutions with a large number of negative decoherence rates.

A very interesting poblem that still remains open is the maximal possible number of eternally negative rates that produce a legitimate solution. Even for the generalized Pauli channels, it is only known that there cannot be more than $d-1$ identical, $(d-1)$-times degenerated rates that are never positive. Another task is a full characterization of P-divisibility for convex combinations of legitimate dynamical maps, even for mixtures of Markovian semigroups.

\section*{Acknowledgements}

K.S. and D.C. were supported by the Polish National Science Centre projects No. 2018/31/N/ST2/00250 and 2018/30/A/ST2/00837, respectively.

\bibliography{C:/Users/cynda/OneDrive/Fizyka/bibliography}
\bibliographystyle{C:/Users/cynda/OneDrive/Fizyka/beztytulow2}

\end{document}